\documentclass[twocolumn,prl,floatfix,citeautoscript,nofootinbib,superscriptaddress]{revtex4}
\usepackage{amsbsy}
\usepackage{latexsym,epsfig,graphicx}
\usepackage{dcolumn}
\usepackage{graphicx}
\usepackage{subfigure}
\usepackage{comment}
\usepackage{color}
\usepackage{bm}
\usepackage{mathrsfs}
\usepackage{amsfonts}
\usepackage{amsmath}
\usepackage{color}
\usepackage{amssymb}
\usepackage{xspace}
\usepackage{epstopdf}
\usepackage{tabularx}
\usepackage{longtable}
\usepackage[colorlinks=true, letterpaper=true, pdfstartview=FitV, linkcolor=blue, citecolor=blue, urlcolor=blue]{hyperref}
\usepackage[normalem]{ulem}

\pdfoutput=1

\begin{document}

\title{Wavelength conversion for single-photon polarization qubits through
continuous variable quantum teleportation}
\author{Xi-Wang Luo}
\affiliation{Department of Physics, The University of Texas at Dallas, Richardson, Texas
75080-3021, USA}
\author{Chuanwei Zhang}
\affiliation{Department of Physics, The University of Texas at Dallas, Richardson, Texas
75080-3021, USA}
\author{Irina Novikova}
\affiliation{Department of Physics, William and Mary, Williamsburg, VA 23187, USA}
\author{Chen Qian}
\affiliation{Department of Computer Science and Engineering, University California Santa
Cruz, Santa Cruz, CA 95064, USA}
\author{Shengwang Du}
\affiliation{Department of Physics, The University of Texas at Dallas, Richardson, Texas
75080-3021, USA}

\begin{abstract}
A quantum internet connects remote quantum processors that need interact and
exchange quantum signals over a long distance through photonic channels.
However, these quantum nodes are usually composed of quantum systems with
emitted photons unsuitable for long-distance transmission. Therefore, 
quantum wavelength conversion to telecom is crucial for long-distance
quantum networks based on optical fiber. Here we propose wavelength
conversion devices for single-photon polarization qubits using continuous
variable quantum teleportation, which can efficiently convert qubits between
near-infrared (780/795 nm suitable for interacting with atomic quantum
nodes) and telecom wavelength (1300-1500 nm suitable for long-distance
transmission). The teleportation uses entangled photon sources (i.e.,
non-degenerate two-mode squeezed state) that can be generated by four-wave
mixing in rubidium atomic vapor cells, with a diamond configuration of
atomic transitions. The entangled fields can be emitted in two orthogonal
polarizations with locked relative phase, making them especially suitable
for interfacing with single-photon polarization qubits. Our work paves the
way for the realization of long-distance quantum networks.
\end{abstract}

\maketitle

%
%

\section{Introduction}

Quantum technologies have been intensively developed in recent years~\cite%
{PRXQuantum.2.017001,PRXQuantum.2.017002,PRXQuantum.2.017003}. In
particular, long-distance quantum communication is crucial for unconditional
security as well as connecting remote quantum processors through quantum
internet~\cite{science.Qinternet,Nature.Qinternet}. A quantum internet is
expected to be more powerful than the simple sum of each quantum nodes,
enabling a number of revolutionary applications such as quantum networks of
clocks~\cite{nphys3000} and distributed quantum computing. Its realization
will rely on the long-distance communication between remote quantum
processors through photonic channels~\cite%
{science.Qinternet,Nature.Qinternet}. However, single-photon qubits emitted
from these quantum nodes (such as atomic ensembles~\cite{nature.Duan},
trapped ions~\cite{RevModPhys.82.1209}), often in the visible or
near-infrared (NIR) regions, are usually unsuitable for long-distance
transmission. Moreover, interconnections between disparate quantum systems
was impossible due to their mismatched emission wavelengths~\cite%
{PRXQuantum.2.017002,science.Qinternet,Nature.Qinternet,Hybrid.quantum}.
Quantum wavelength conversion (QWC)~\cite{QWC}, which enables the spectral
translation of a photon to targeted wavelength without disturbing its
quantum properties, is a solution to these issues.

\begin{figure}[tb]
\includegraphics[width=1.0\linewidth]{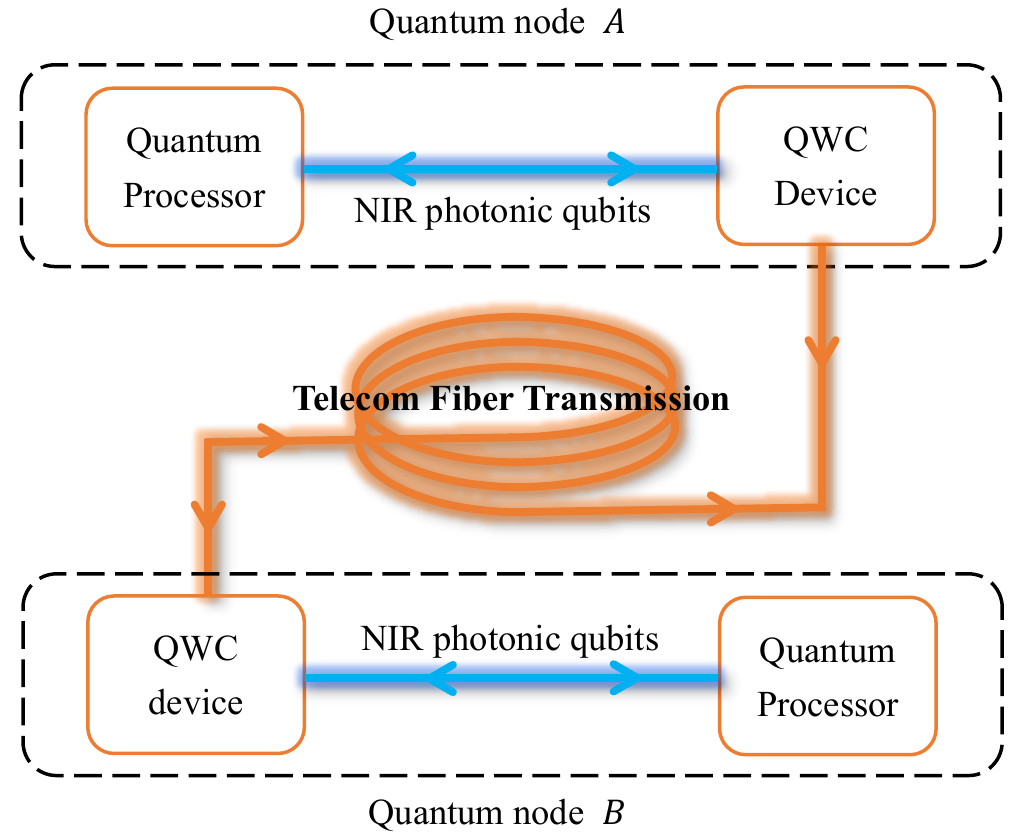}
\caption{Illustration of long-distance communication between remote quantum
nodes.}
\label{fig:QI}
\end{figure}

QWC between the NIR and telecom bands~\cite{nature04009} is one of the most
important wavelength conversions, since telecom band photons remain the
information carrier of choice for long-distance transmission based on
optical fiber, while the NIR photons not only interact with atomic quantum
systems but also fall into the working band of many high-performance
single-photon detectors~\cite{nphoton.2009.230} and quantum memories~\cite%
{RevModPhys.83.33}. For long-distance communication between remote quantum
nodes, one first uses the QWC device to convert the single-photon qubit
emitted by quantum node $A$ from NIR to telecom wavelength; then send it to
quantum node $B$ through fiber transmission; and finally convert its
wavelength back to NIR band for interacting with the quantum processor (as
illustrated in Fig.~\ref{fig:QI}). QWC between NIR and telecom bands has
been demonstrated for application to a wide range of quantum systems (e.g.,
trapped ions and rubidium gas)~\cite%
{ncomms1544,nphys1773,PhysRevLett.105.260502,PhysRevLett.120.203601,ncomms1998,nature.578.240,nature24468, NC1997}%
. However, these QWC devices are usually based on three- or four-wave
nonlinear optical mixing, which still has difficulty about noise photon and
conversion efficiency ($30\% \sim 70\%$) to realize photonic interface. 

In this paper, we propose wavelength conversion devices for single-photon
polarization qubits using continuous variable (CV) quantum teleportation~%
\cite%
{PhysRevA.49.1473,PhysRevLett.80.869,science.282.5389,PhysRevA.65.012313}.
The entangled sources of the CV teleporters are non-degenerate two-mode
squeezed vacuum (TMSV) states. By performing a join homodyne measurement of
a single photon and one of the entangled fields at the same NIR (telecom)
wavelength, we can then teleport the single-photon qubit into the telecom
(NIR) band, by imprinting the measured joint quadratures into the other
entangled field at the shifted wavelength. The use of a hybrid technique
involving CV teleportation of a discrete-variable (i.e., polarization
qubits) allows deterministic wavelength conversion of photonic qubits with
nearly 100\% efficiency.

Since homodyne measurements~\cite{Wiseman2010quantum} require using light of
known polarization, therefore, two polarizations need be teleported
parallelly and each polarization component of the qubit requires a TMSV
entangled state in the teleportation. Additional phase locking between the
two polarization teleporters (i.e., the two TMSV entangled states) is
necessary to avoid any phase errors in the teleportation, which is still
challenging even for ordinary CV teleportation without wavelength
conversion, where the degenerate TMSV entangled source is generated by
mixing two identical single-mode squeezed states at a 50:50 beam splitter. 

To realized QWC, we need to use \textit{non-degenerate} TMSV states which
cannot be generated by mixing single-mode squeezed states. We propose to
generate the entangled source using a four-wave mixing (FWM) process in a
hot Rubidium vapor cell with a diamond configuration of atomic transitions~%
\cite{PhysRevLett.96.093604,PhysRevLett.111.123602,NJP.17,OSAC.2.002260}, in
which the wavelength of one of the fields matches the single photon emitted
from the atomic quantum nodes, and the other falls into the telecom-band
optical fields, suitable for low loss fiber transmission. Because of the
symmetry, the TMSV states can be emitted in two orthogonal polarizations
with fixed relative phase, making them especially suitable for interfacing
with polarization single photon qubits. Our approach provides an attractive
alternative to a nonlinear crystal-based wavelength conversion that
typically has low efficiency and requires tremendous technical care. In
principle, the CV teleportation based wavelength conversion can be
generalized to a wide range of frequencies as long as suitable two-color
squeezed states can be generated using non-linear optical mixing.

\section{QWC through CV teleportation}

Quantum teleportation~\cite{PhysRevLett.70.1895} is a technique for
transferring quantum information from a sender at one location to a receiver
some distance away, by sending only classical information using a shared
entangled state as a resource. It has become one of the key elements of
advanced and practical quantum information protocols. Originally quantum
teleportation was proposed for discrete variables qubit, and much progress
has been made in demonstrating quantum teleportation of photonic qubits.
However, most of these schemes share one fundamental restriction: they
require an unambiguous two-qubit Bell-state measurement which is always
probabilistic (with a success probability 50\%) when linear optics is used~%
\cite{RevModPhys.84.777}.

The concept was generalized to CV teleportation~\cite{PhysRevA.49.1473},
which relies on the quadrature-entangled states (i.e., two-mode squeezed
states)~\cite{arXiv.1401.4118} and the measurements in the quadrature bases
using linear optics and homodyne detection, leading to deterministic
teleportation without post-selection. So far, the CV teleportation has been
used to unconditionally teleport quantum states such as nonclassical CV
state and time-bin qubits with fixed polarization~\cite%
{PhysRevA.67.032302,PhysRevLett.93.250503,PhysRevLett.94.220502,PhysRevLett.99.110503,science.1201034,nature12366}%
. 
Since homodyne measurements require light of known polarization, one need a
pair of such entangled states with locked relative phase to parallelly
teleport the two spin components of the qubits.

We consider non-degenerate two-mode squeezed states generated by the
following Hamiltonian 
\begin{equation}
H_{\text{sq}}=\sum_{s=H,V}[i\xi _{s}\hat{a}_{A,s}^{\dag }\hat{a}_{B,s}^{\dag
}+h.c.],  \label{Ham}
\end{equation}%
which leads to squeezing between modes $A$ and $B$ for both polarizations $H$
and $V$. Here $\hat{a}_{A,B}^{\dag }$ are the photon creation operators of
modes $A$ and $B$, with corresponding frequencies $\omega _{A,B}$ falling in
the NIR and telecom bands respectively. The entangled source $|\text{ES}%
\rangle =e^{-iH_{\text{sq}}\tau }|\text{vac}\rangle $ is generated by
evolving the vacuum under the Hamiltonian $H_{\text{sq}}$ for certain time
period $\tau $. We obtain a pair of two-mode squeezed vacuum $|\text{ES}%
\rangle _{\Omega }=\prod_{s}\hat{S}_{s}(r_{s})|\text{vac}\rangle $, with $%
\hat{S}_{s}(r)=e^{r\hat{a}_{A,s}^{\dag }\hat{a}_{B,s}^{\dag }-h.c.}$ the
two-mode squeezing operator~\cite{arXiv.1401.4118,walls2008quantum} for
polarization $s$ and $r_{s}=\xi _{s}\tau $ the squeezing factor. We assume $%
r_{H}=r_{V}=r$ are real, therefore, a pair of in-phase non-degenerate TMSV
states are generated coherently.

\begin{figure*}[tb]
\includegraphics[width=1.0\linewidth]{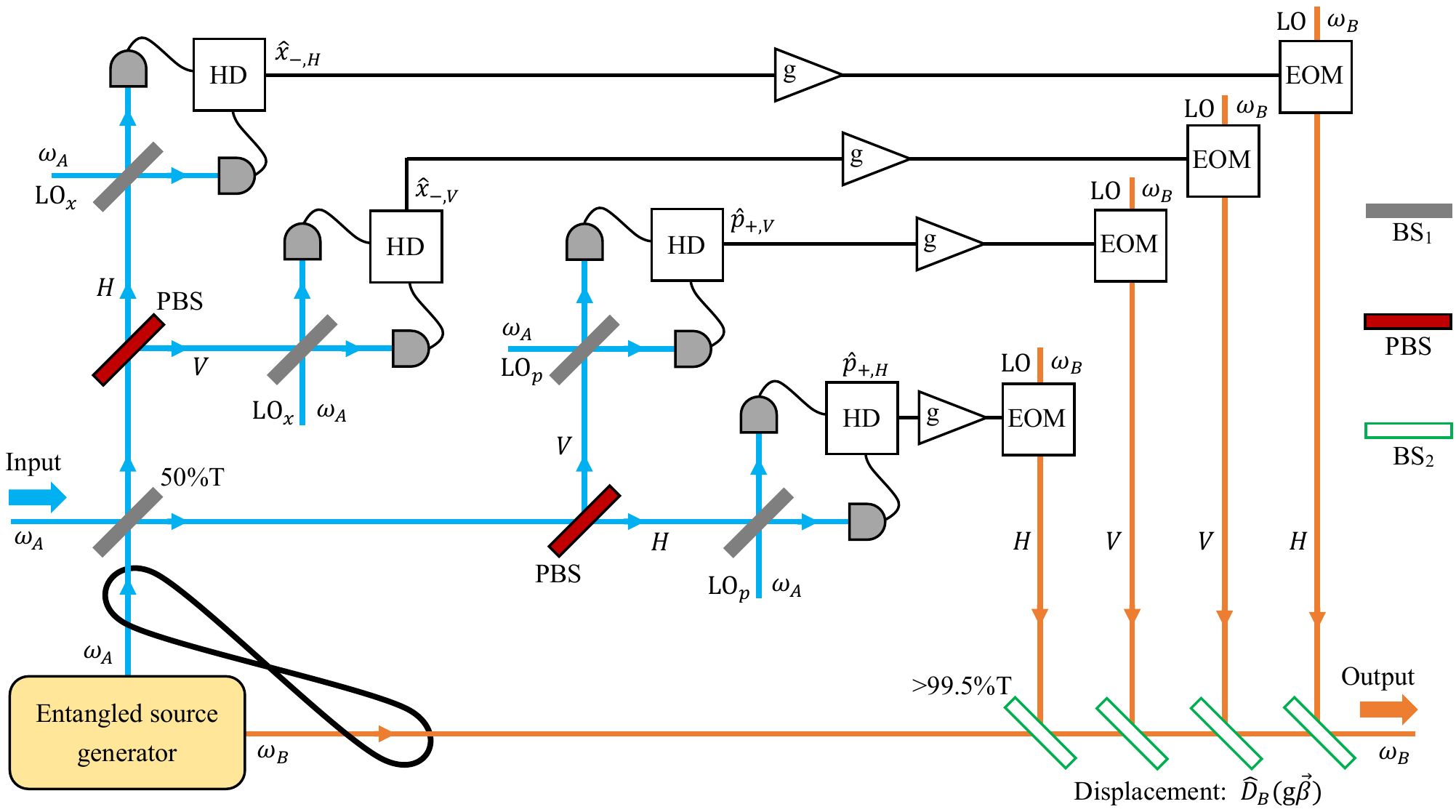}
\caption{Schematic illustration of the quantum wavelength conversion device
based on CV quantum teleportation. BS$_{1}$ and BS$_{2}$ represent the beam
splitters with 50\% and $>$99.5\% transmittance, respectively; PBS, HD and
EOM represent polarizing beam splitter, homodyne detector and electro-optic
modulator. LO$_{x}$ and LO$_{p}$ are local oscillators to measure $x$ and $p$
quadratures, respectively. For the homodyne detection (displacement), the
frequencies of local oscillators are $\protect\omega _{A}$ ($\protect\omega %
_{B}$).}
\label{fig:QWC}
\end{figure*}

In the photon number basis, the entangled source can be written as~\cite%
{PhysRevA.65.012313} 
\begin{eqnarray}
|\text{ES}\rangle &=&(1-q^{2})\sum_{n,m}q^{n+m}|n;m\rangle _{A}|n;m\rangle
_{B}  \notag \\
&=&(1-q^{2})\sum_{n,N}q^{N}|n;N-n\rangle _{A}|n;N-n\rangle _{B}
\end{eqnarray}%
where $|n;N-n\rangle $ represents $n$ photons in $H$ polarization and $N-n$
photons in $V$ polarization, and $q=\tanh (r)$. We want to teleport a
single-photon polarization qubit $|\psi \rangle _{\text{in}%
}=c_{1}|1;0\rangle _{\text{in}}\ +c_{2}|0;1\rangle _{\text{in}}$ and convert
its frequency from $\omega _{A}$ to $\omega _{B}$.

The quantum wavelength conversion device based on CV teleportation process
is schematically illustrated in Fig.~\ref{fig:QWC}. First, we combine the
input mode with mode $A$ at a half beam splitter and make the balanced
homodyne detection of mode $\hat{a}_{\pm ,s}=\frac{\hat{a}_{\text{in},s}\pm 
\hat{a}_{A,s}}{\sqrt{2}}$ with $\hat{a}_{\text{in},s}$ the input field
operator for polarization $s$. We use a local oscillator with frequency $%
\omega _{A}$ to measure $\hat{x}_{-,s}=(\hat{a}_{-,s}+\hat{a}_{-,s}^{\dag })/%
\sqrt{2}$ and $\hat{p}_{+,s}=i(\hat{a}_{+,s}^{\dag }-\hat{a}_{+,s})/\sqrt{2}$%
, which generates the control signal $\Vec{\beta}=(\beta _{H},\beta _{V})$,
where $\beta _{s}=\langle \hat{x}_{-,s}+i\hat{p}_{+,s}\rangle $ and $\langle
\cdot \rangle $ corresponds to the detection projection. The homodyne
measurement projects the state of input and $A$ modes to the state 
\begin{equation}
|\Vec{\beta}\rangle _{\text{in},A}=\frac{1}{\pi }\sum_{n,m}\hat{D}_{\text{in}%
}(\Vec{\beta})|n;m\rangle _{\text{in}}|n;m\rangle _{A}
\end{equation}%
with displacement operator $\hat{D}_{\text{in}}(\Vec{\beta}%
)=\prod_{s}e^{\beta _{s}{\hat{a}_{\text{in},s}}^{\dagger }-\beta _{s}^{\ast }%
{\hat{a}}_{\text{in},s}}$. Then we perform the displacement $\text{g}\Vec{%
\beta}$ on mode $B$ using a local oscillator with frequency $\omega _{B}$,
and accomplish the teleportation. Here $\text{g}$ is the gain factor and the
output state (unnormalized) in mode $B$ becomes 
\begin{eqnarray}
|\psi \rangle _{\text{out}} &=&_{\text{in},A}\langle \Vec{\beta}|\hat{D}_{B}(%
\text{g}\Vec{\beta})|\text{ES}\rangle |\psi \rangle _{\text{in}}  \notag \\
&=&\hat{T}_{q}(\Vec{\beta})|\psi \rangle _{\text{in}},
\end{eqnarray}%
where 
\begin{equation*}
\hat{T}_{q}(\Vec{\beta})=\frac{1-q^{2}}{\pi }\sum_{n,m}q^{n+m}\hat{D}_{B}(%
\text{g}\Vec{\beta})|n;m\rangle _{B\text{ in}}\langle n;m|\hat{D}_{\text{in}%
}(-\Vec{\beta})
\end{equation*}%
is the transfer operator. In the strong squeezing limit, we have $\hat{T}%
_{q}(\Vec{\beta})\propto \sum_{n,m}|n;m\rangle _{B\text{ in}}\langle n;m|$
for unit gain $\text{g}=1$, and the normalized output state reads 
\begin{equation}
|\psi \rangle _{\text{out}}=c_{1}|1;0\rangle _{B}+c_{2}|0;1\rangle _{B},
\end{equation}%
which is identical to the input state, except that the frequency of the
qubit is shifted from $\omega _{A}$ to $\omega _{B}$. The teleportation
process corresponds to one-photon teleportation in $H$ ($V$) polarization
and a vacuum teleportation in $V$ ($H$) polarization. Similarly, we can
convert the wavelength of the photonic qubit from $\omega _{B}$ to $\omega
_{A}$.

To generate the displacement on mode $B$, we can use electro-optical
modulators (EOMs) to modulate the local oscillator (the local oscillator has
frequency $\omega_B$) according to the homodyne detection signals, and
combine these phase modulated beams with mode $B$ at a beam splitter of $>$%
99.5\% transmittance (as shown in Fig.~\ref{fig:QWC}). Moreover, we want to
mention that the ideal two-mode squeezing requires infinite energy, and
ideal CV entangled source is physically unattainable; thus, the
teleportation fidelity is generally limited by the squeezing factor. In this
case, non-unit gain conditions are useful~\cite%
{nature12366,PhysRevA.64.040301,PhysRevLett.83.2095}. With proper choice of
the gain factor $\text{g}=q=\tanh(r)$, no additional photons would be
created in the output, and the teleported single photon qubit remains
undisturbed regardless of the squeezing level, with a success probability $%
q^2$~\cite{nature12366,PhysRevA.64.040301,PhysRevLett.83.2095}. Therefore,
the wavelength conversion efficient (i.e., $q^2$) can be nearly 100\% for
sufficiently strong squeezing $\tanh(r)\rightarrow 1$.

\section{Entangled source from FWM}

The key element of our scheme is to generate a pair of non-degenerate TMSV
states, where the wavelength of one mode in the TMSV state matches photons
emitted from the quantum nodes, and the other mode falls into the telecom
band. In this section, we show how to generate such non-degenerate two-mode
squeezing through four-wave mixing.

As we discussed previously, homodyne measurements require using light of
known polarization, therefore, two polarizations need to be teleported
parallelly and each polarization component of the qubit requires a TMSV
entangled state. The relative phase between the two TMSV states should be
locked. Suppose there is a random and unknown relative phase between the two
polarizing TMSV states (i.e., we replace $\hat{a}_{A,V}$ and $\hat{a}_{B,V}$
by $e^{-i\phi _{A}}\hat{a}_{A,V}$ and $e^{-i\phi _{B}}\hat{a}_{A,V}$ in Eq.~%
\ref{Ham}), then the entangled state becomes 
\begin{equation*}
|\text{ES}(\phi )\rangle =(1-q^{2})\sum_{n,m}q^{n+m}e^{im\phi }|n;m\rangle
_{A}|n;m\rangle _{B}
\end{equation*}%
with $\phi =\phi _{A}+\phi _{B}$ a random and unknown phase. The final
output state becomes 
\begin{equation}
|\psi (\phi )\rangle _{\text{out}}=c_{1}|1;0\rangle _{B}+c_{2}e^{i\phi
}|0;1\rangle _{B}
\end{equation}%
with a phase error. Therefore, even the ordinary CV teleportation (without
wavelength conversion) of a single-photon polarization qubit is still
challenging and has not been demonstrated experimentally, since a TMSV state
generated by mixing two identical single-mode squeezed states usually has an
unknown relatively phase with respect to other TMSV states generated in the
same way.

\begin{figure}[tb]
\includegraphics[width=1.0\linewidth]{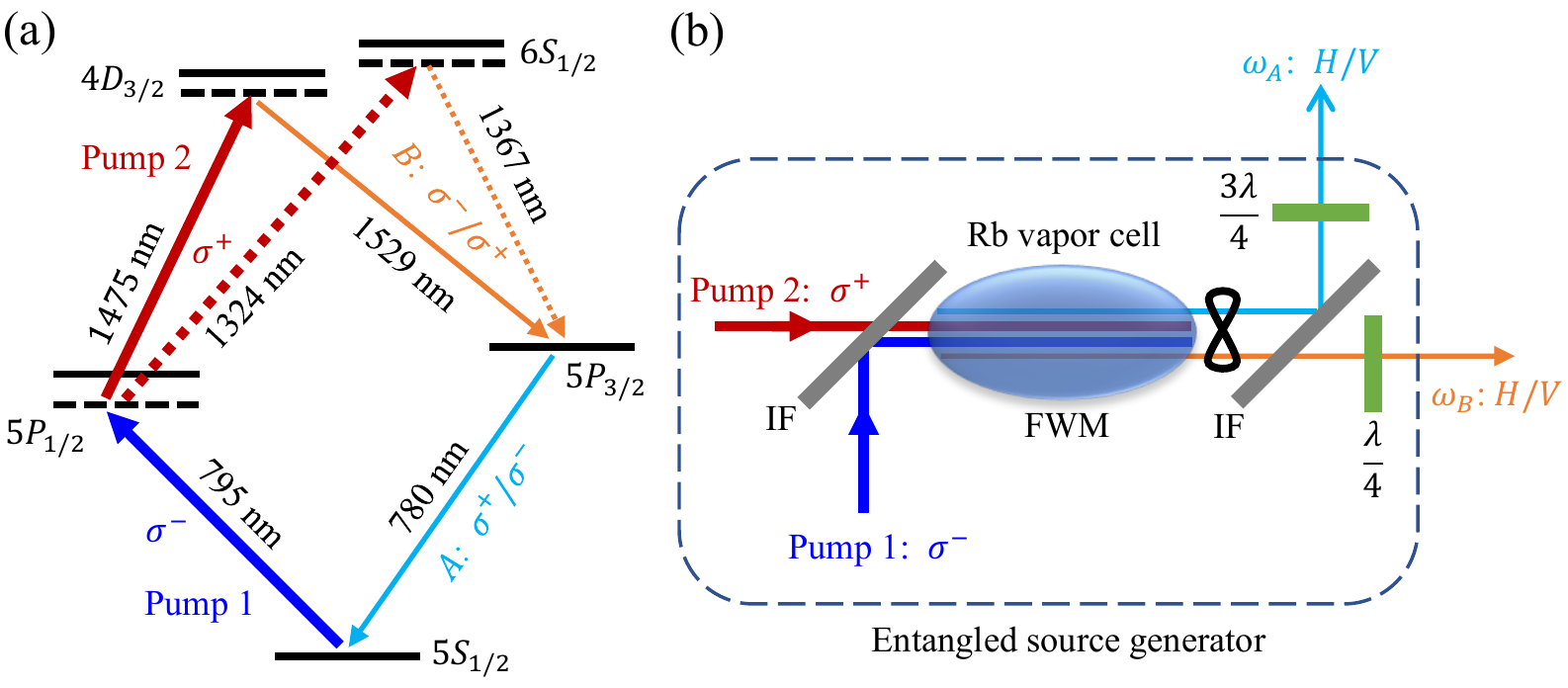}
\caption{Illustration of the FWM process to generate two-mode squeezing. (a)
Level scheme for the FWM in $^{87}$Rb. (b) Schematic of the proposed
entangled source generator. IF represents the interference filter. The $%
\frac{3\protect\lambda }{4}$ and $\frac{\protect\lambda }{4}$ waveplates are
used to convert $\protect\sigma ^{\pm }$ to $H/V$ polarizations.}
\label{fig:FWM}
\end{figure}

To realized QWC through CV teleportation, we need to use non-degenerate TMSV
states which cannot be generated by mixing single-mode squeezed states. Here
we consider the four-wave mixing in a hot Rubidium vapor cell~\cite%
{PhysRevLett.96.093604,PhysRevLett.111.123602,NJP.17,OSAC.2.002260,OL.44.005921,OPTICA.4.000272}%
, with a diamond configuration of atomic transitions, as shown in Fig.~\ref%
{fig:FWM}a. The ground state $5S_{1/2}$ is coupled to the excited state $%
4D_{3/2}$ through a two-photon process, with intermediate states $5P_{1/2}$
or $5P_{3/2}$. The setup is schematically shown in Fig.~\ref{fig:FWM}b. An
ensemble of $^{87}$Rb atoms is trapped with a magneto-optical trap, all
atoms are initialized to ground state $5S_{1/2}$. In the FWM process, two
pump lasers (with NIR wavelength $\lambda _{1}=795$ nm and telecom $\lambda
_{2}=1475$ nm, respectively) are used to excite atoms to state $4D_{3/2}$
with intermediate state $5P_{1/2}$. Then state $4D_{3/2}$ decays back to $%
5S_{1/2}$ through parametric down conversion with intermediate state $%
5P_{3/2}$, followed by the generation of photon pairs (with telecom
wavelength $\lambda _{B}=1529$ nm and NIR wavelength $\lambda _{A}=780$ nm,
respectively). The Hamiltonian of the FWM process can be written as 
\begin{equation}
H_{\text{FWM}}=\chi ^{\text{(3)}}\ \hat{a}_{1}\hat{a}_{2}\hat{a}_{A}^{\dag }%
\hat{a}_{B}^{\dag }+h.c.,  \label{Ham2}
\end{equation}%
where $\chi ^{\text{(3)}}$ is the third-order nonlinear coefficient and $%
a_{j}$ is the annihilation operator for optical field $\lambda _{j}$. We can
replace the field operator $\hat{a}_{1,2}$ of the pump lasers by the
coherent amplitudes $\hat{\alpha}_{1,2}$, then $H_{\text{FWM}}\rightarrow
i\xi \hat{a}_{A}^{\dag }\hat{a}_{B}^{\dag }+h.c.$ with $\xi =-i\chi ^{\text{%
(3)}}\ \hat{\alpha}_{1}\hat{\alpha}_{2}$ to be real for proper gauge choice.
Energy conservation requires that the generated photon pairs in mode $A$ and 
$B$ have frequencies $\omega _{A}$ and $\omega _{B}$, respectively, where $%
\omega _{j}=\frac{2\pi c}{\lambda _{j}}$ satisfies $\omega _{A}+\omega
_{B}=\omega _{1}+\omega _{2}$ and $c$ is the speed of light.


Now we examine the polarizations of the two-mode squeezing fields generated
by FWM. We consider that the four fields are in a co-propagating geometry
inside the atomic cloud, which satisfies the phase matching, as shown in
Fig.~\ref{fig:FWM}b. With the quantization axis along the beam propagation
direction of all modes, we drive transitions with $\Delta m_{F}=\pm 1$ using
two pump beams that are orthogonally circularly polarized. In the coherent
parametric down conversion process, the final quantum state of the atoms
remains the same as the initial state. Furthermore, rotational symmetry of
the atomic cloud along beam propagation direction implies angular momentum
conservation. The angular momentum selection rules limit the polarizations
of the generated photon pairs, and the Hamiltonian becomes 
\begin{equation}
H_{\text{FWM}}=i\xi (\hat{a}_{A,\sigma ^{+}}^{\dag }\hat{a}_{B,\sigma
^{-}}^{\dag }+s\hat{a}_{A,\sigma ^{-}}^{\dag }\hat{a}_{B,\sigma ^{+}}^{\dag
}+h.c.),  \label{Ham3}
\end{equation}%
where $\sigma ^{\pm }$ denote the left and right circular polarizations,
respectively, and $s$ is determined by the Clebsh-Gordan coefficients~\cite%
{NJP.17}. We can choose the initial state as $5S_{1/2},F=2,m_{F}=0$, which
leads to $s=1$. Therefore, with proper local operations on the entangled
fields $\hat{a}_{A,\sigma ^{\pm }}\rightarrow \hat{a}_{A,HV}$ and $\hat{a}%
_{B,\sigma ^{\mp }}\rightarrow \hat{a}_{B,HV}$, $H_{\text{FWM}}$ reduces to $%
H_{\text{sq}}$ and the FWM process generates two-mode squeezing for both
polarizations with locked and known relative phase, which makes them
especially suitable for teleporting single-photon polarization qubits.

We want to point out that, the parametric down conversion process may have
multi decay paths through different hyperfine levels of $5P_{3/2}$, we can
use additional filters to select one decay path (e.g., $5P_{3/2},F=3$). We
can also use $\lambda_1=780$, $\lambda_2=1529$ for the pump beams with $%
5P_{3/2}$ the corresponding intermediate state, then we obtain the two-mode
squeezing with wavelength $\lambda_{A}=795$ and $\lambda_{B}=1475$.
Moreover, we can choose a different atomic level such as $6S_{1/2}$ to be
the high exited state (see Fig.~\ref{fig:FWM}a), which allows us to generate
telecom fields with wavelength $1324$ nm and $1367$ nm.

\section{Finite qubit bandwidth}

We have assumed a single-frequency photonic qubit in the discussion above,
while in practice, the photonic qubit always has a finite frequency
bandwidth. In the Heisenberg picture, the balanced homodyne detection
corresponds to projection measurement $\beta _{s}(t)=\langle e^{i\omega
_{A}t}\hat{a}_{\text{in},s}(t)-e^{-i\omega _{A}t}\hat{a}_{A,s}^{\dag
}(t)\rangle $. After performing the displacement on mode $B$, the output
fields are 
\begin{equation}
\hat{a}_{\text{out},s}(t)=\hat{a}_{B,s}(t)+e^{-i\omega _{-}t}\hat{a}_{\text{%
in},s}(t)-e^{-i\omega _{+}t}\hat{a}_{A,s}^{\dag }(t),
\end{equation}%
with $\omega _{\pm }=\omega _{B}\pm \omega _{A}$. In the frequency domain,
we have 
\begin{eqnarray}
\hat{a}_{\text{out},s}(\omega _{B}+\Omega ) &=&\hat{a}_{\text{in},s}(\omega
_{A}+\Omega )  \label{eq:Heisenberg} \\
&+&\hat{a}_{B,s}(\omega _{B}+\Omega )-\hat{a}_{A,s}^{\dag }(\omega
_{A}-\Omega ).  \notag
\end{eqnarray}%
This means that, we need two-mode squeezing between frequency component $%
\omega _{A}-\Omega $ in mode $A$ and the frequency component $\omega
_{B}+\Omega $ in mode $B$ to convert a single-frequency qubit from $\omega
_{A}+\Omega $ to $\omega _{B}+\Omega $.

Meanwhile, we can take into account the finite squeezing bandwidth of the
FWM process, and the Hamiltonian reads 
\begin{equation}
H_{\text{FWM}}=i\xi \sum_{s,\Omega }[\hat{a}_{A,s}^{\dag }(\omega
_{A}-\Omega )\hat{a}_{B,s}^{\dag }(\omega _{B}+\Omega )+h.c.]
\label{Ham_band}
\end{equation}%
due to energy conservation, with $\Omega $ taking values within the
squeezing bandwidth. The Hamiltonian in Eq.~\ref{Ham_band} leads to exactly
the desired squeezing. In the Heisenberg picture, the TMSV state satisfies $[%
\hat{a}_{B,s}(\omega _{B}+\Omega )-\hat{a}_{A,s}^{\dag }(\omega _{A}-\Omega
)]_{\text{TMSV}}=e^{-r}[\hat{a}_{B,s}(\omega _{B}+\Omega )-\hat{a}%
_{A,s}^{\dag }(\omega _{A}-\Omega )]_{\text{vac}}$. In the strong squeezing
limit, the output field Eq.~\ref{eq:Heisenberg} becomes exactly the same as
the input filed $\hat{a}_{\text{out},s}(\omega _{B}+\Omega )=\hat{a}_{\text{%
in},s}(\omega _{A}+\Omega )$, except the frequency is shifted by $\omega
_{B}-\omega _{A}$. Therefore, our proposal works as long as the squeezing
bandwidth is sufficiently wide to cover the qubit bandwidth.

\section{Discussion and conclusion}

We have shown how to realize the wavelength conversion for single-photon
polarization qubits through CV teleportation, and to generalize the two-mode
squeezed entangled source using FWM. For long-distance communication between
remote quantum nodes, we can use the procedure as illustrated in Fig.~\ref%
{fig:QI}. Alternatively, we can first send telecom mode of the entangled
source from node $A$ to node $B$; then teleport the photonic qubit from node 
$A$ to $B$ with wavelength conversion; and finally convert the qubit back to
NIR wavelength at node $B$. For the later approach, additional photon loss
may be introduced to the entangled source during the transmission, which
will reduce the teleportation fidelity~\cite{PRJ.7.0000A7}.

In summary, we proposed a QWC device for single-photon polarization qubits
using continuous variable quantum teleportation. We considered a four-wave
mixing process in rubidium atomic vapor cell with a diamond configuration of
atomic transitions to generate the entangled source, which corresponds to a
pair of in-phase non-degenerate TMSV states. One of the entangled fields has
wavelength 780/795 nm and the other has wavelength 1300-1500 nm, allowing
efficiently wavelength conversion of single-photon qubits between NIR
(suitable for interacting with atomic quantum nodes) and telecom-wavelength
(suitable for long-distance transmission). Moreover, it is possible to
generate the wavelength conversion scheme to a wide range of frequencies as
long as the corresponding two-color squeezed states can be generated by
suitable non-linear optical mixing. Our work provides an attractive
alternative to a nonlinear crystal-based wavelength conversion that still
has difficulty about noise photon and conversion efficiency.


\section{Acknowledgement}

X.W.L. and C.Z. are supported by AFOSR (FA9550-20-1-0220), NSF
(PHY-1806227), and ARO (W911NF-17-1-0128). S.D. acknowledges the Texas STARs
and Start-Up fundings from The University of Texas at Dallas. I.N. is
supported by AFOSR (FA9550-19-1-0066). C.Q. was supported by NSF
(CNS-1750704 and CPS-1932447).

\end{document}